\documentclass[dvips]{acta}
\usepackage{supertabular,lscape,epsfig}
\usepackage{amssymb}
\usepackage{amsmath}
\usepackage{float}
\mathchardef\mhyphen="2D
\usepackage{placeins}
\DeclareSymbolFont{ppa}{OT1}{ppl}{m}{it}
\DeclareMathSymbol{\vv}{\mathalpha}{ppa}{'166}

\SetPages{363}{378}

\SetVol{67}{2017}

\begin{document}
\newcommand\pvalue{\mathop{p\mhyphen {\rm value}}}
\newcommand{\TabApp}[2]{\begin{center}\parbox[t]{#1}{\centerline{
  {\bf Appendix}}
  \vskip2mm
  \centerline{\small {\spaceskip 2pt plus 1pt minus 1pt T a b l e}
  \refstepcounter{table}\thetable}
  \vskip2mm
  \centerline{\footnotesize #2}}
  \vskip3mm
\end{center}}

\newcommand{\TabCapp}[2]{\begin{center}\parbox[t]{#1}{\centerline{
  \small {\spaceskip 2pt plus 1pt minus 1pt T a b l e}
  \refstepcounter{table}\thetable}
  \vskip2mm
  \centerline{\footnotesize #2}}
  \vskip3mm
\end{center}}

\newcommand{\TTabCap}[3]{\begin{center}\parbox[t]{#1}{\centerline{
  \small {\spaceskip 2pt plus 1pt minus 1pt T a b l e}
  \refstepcounter{table}\thetable}
  \vskip2mm
  \centerline{\footnotesize #2}
  \centerline{\footnotesize #3}}
  \vskip1mm
\end{center}}

\newcommand{\MakeTableApp}[4]{\begin{table}[p]\TabApp{#2}{#3}
  \begin{center} \TableFont \begin{tabular}{#1} #4 
  \end{tabular}\end{center}\end{table}}

\newcommand{\MakeTableSepp}[4]{\begin{table}[p]\TabCapp{#2}{#3}
  \begin{center} \TableFont \begin{tabular}{#1} #4 
  \end{tabular}\end{center}\end{table}}

\newcommand{\MakeTableee}[4]{\begin{table}[htb]\TabCapp{#2}{#3}
  \begin{center} \TableFont \begin{tabular}{#1} #4
  \end{tabular}\end{center}\end{table}}

\newcommand{\MakeTablee}[5]{\begin{table}[htb]\TTabCap{#2}{#3}{#4}
  \begin{center} \TableFont \begin{tabular}{#1} #5 
  \end{tabular}\end{center}\end{table}}

\newcommand{\MakeTableH}[4]{\begin{table}[H]\TabCap{#2}{#3}
  \begin{center} \TableFont \begin{tabular}{#1} #4 
  \end{tabular}\end{center}\end{table}}

\newcommand{\MakeTableHH}[4]{\begin{table}[H]\TabCapp{#2}{#3}
  \begin{center} \TableFont \begin{tabular}{#1} #4 
  \end{tabular}\end{center}\end{table}}

\newfont{\bb}{ptmbi8t at 12pt}
\newfont{\bbb}{cmbxti10}
\newfont{\bbbb}{cmbxti10 at 9pt}
\newcommand{\uprule}{\rule{0pt}{2.5ex}}
\newcommand{\douprule}{\rule[-2ex]{0pt}{4.5ex}}
\newcommand{\dorule}{\rule[-2ex]{0pt}{2ex}}
\begin{Titlepage}
\Title{OGLE Collection of Star Clusters. \\
New Objects in the Magellanic Bridge and the Outskirts of the Small Magellanic Cloud}

\Author{M.~~S~i~t~e~k$^1$, ~~M.\,K.~~S~z~y~m~a~ñ~s~k~i$^1$,
~~A.~~U~d~a~l~s~k~i$^1$,~~D.\,M.~~S~k~o~w~r~o~n$^1$,
~~Z.~~K~o~s~t~r~z~e~w~a~-~R~u~t~k~o~w~s~k~a$^{2,3}$, ~~J.~~S~k~o~w~r~o~n$^1$,
~~P.~~K~a~r~c~z~m~a~r~e~k$^1$, ~~M.~~C~i~e~{\'s}~l~a~r$^1$,
~~£.~~W~y~r~z~y~k~o~w~s~k~i$^1$, ~~S.~~K~o~z~³~o~w~s~k~i$^1$,
~~P.~~P~i~e~t~r~u~k~o~w~i~c~z$^1$, ~~I.~~S~o~s~z~y~ñ~s~k~i$^1$,
~~P.~~M~r~ó~z$^1$, ~~M.~~P~a~w~l~a~k$^{1,4}$, ~~R.~~P~o~l~e~s~k~i$^{1,5}$
~~and~~K.~~U~l~a~c~z~y~k$^{1,6}$}
{$^1$Warsaw University Observatory, Al~Ujazdowskie~4, 00-478~Warszawa, Poland\\
e-mail:(msitek,msz)@astrouw.edu.pl\\
$^2$SRON Netherlands Institute for Space Research, Sorbonnelaan 2, 3584 CA Utrecht,\\ the Netherlands\\
$^3$Department of Astrophysics/IMAPP, Radboud University Nijmegen, P.O. Box 9010, 6500 GL Nijmegen, the Netherlands\\
$^4$Institute of Theoretical Physics, Faculty of Mathematics and Physics, Charles University in Prague, Czech Republic\\
$^5$Department of Astronomy, Ohio State University, 140 W. 18th Ave., Columbus,\\ OH 43210, USA\\
$^6$Department of Physics, University of Warwick, Gibbet Hill Road, Coventry,\\ CV4 7AL, UK
}
\end{Titlepage}

\Abstract{The Magellanic System (MS) encompasses the nearest neighbors of
  the Milky Way, the Large (LMC) and Small (SMC) Magellanic Clouds, and the
  Magellanic Bridge (MBR). This system contains a diverse sample of star
  clusters. Their parameters, such as the spatial distribution, chemical
  composition and age distribution yield important information about the
  formation scenario of the whole Magellanic System. Using deep photometric
  maps compiled in the fourth phase of the Optical Gravitational Lensing
  Experiment (OGLE-IV) we present the most complete catalog of star
  clusters in the Magellanic System ever constructed from homogeneous, long
  time-scale photometric data.  In this second paper of the series, we show
  the collection of star clusters found in the area of about 360 square
  degrees in the MBR and in the outer regions of the SMC. Our sample
  contains 198 visually identified star cluster candidates, 75 of which
  were not listed in any of the previously published catalogs. The new
  discoveries are mainly young small open clusters or clusters similar to
  associations.}{Catalogs -- Galaxies: star clusters: general -- Surveys}

\Section{Introduction}
\vspace*{3pt}
The Magellanic System (MS) provides an excellent astrophysical laboratory
for studying the structure and evolution of stellar systems (Skowron \etal
2014, Piatti \etal 2015, Jacyszyn-Dobrzeniecka \etal 2016). Star clusters
are one of the tools for such studies. However, a complete collection of
star clusters is needed to conduct such a research derived from homogeneous
observational data, preferably from a single photometric survey. For more
details of the scientific rationale of this research, see the Introduction
in Sitek \etal (2016, hereafter referred to as Paper~I).

To date, the largest catalog of extended objects in the Magellanic System
was prepared by Bica \etal (2008) as a compilation of all the previously
published catalogs. The important contribution to this sample was
taken from the OGLE-II star clusters catalogs (Pietrzyñski
\etal 1998, 1999). These catalogs, however, covered only the central parts
of the LMC and SMC: 5.8 and 2.5 square degrees (Udalski \etal 1997)
respectively -- only 1.5--2\% of the area observed toward these regions
during the current OGLE-IV phase (Udalski, Szymañski and Szymañski
2015). The Magellanic Bridge has never been systematically observed at such
scale before, both in terms of the area, time range and cadence. The MBR
coverage in the OGLE-IV is 185 square degrees.

This paper presents the second part of the star cluster collection based on
the OGLE-IV data. The central part of the SMC has already been observed or
analyzed by many other projects (Piatti and Bica 2012, Piatti 2016). Thus,
we decided to start our exploration from the outer parts of the galaxy. We
have also analyzed the whole area of the MBR covered by the OGLE-IV survey.

\vspace*{7pt}
\Section{Observations and Data Reduction}
\vspace*{3pt}
The photometric data of the SMC and MBR fields analyzed in this paper are
based on the images gathered during the first five years (2010--2015) of
the fourth phase of the OGLE project (Udalski, Szymañski and Szymañski
2015).

We have used the ``deep photometric maps'' -- catalogs of all objects
detected on the deep images of all observed fields. For details we refer
the reader to Paper~I. For all 120 observed SMC fields
(which were analyzed in this paper) and 132 MBR fields, the number of the
stacked images used for the deep images is between 55 and 100 (86 on
average), depending on the overall number of good seeing individual images
in the {\it I}-band available for any given field. For the {\it V}-band,
which is observed less frequently, the deep images were constructed from 2
to 100 individual images with the mean value of 13. For comparison, the
reference images for 41 OGLE-III SMC fields (Udalski \etal 2008) were
constructed using 4--15 images.
\begin{figure}[htb]
\centerline{\includegraphics[width=6cm]{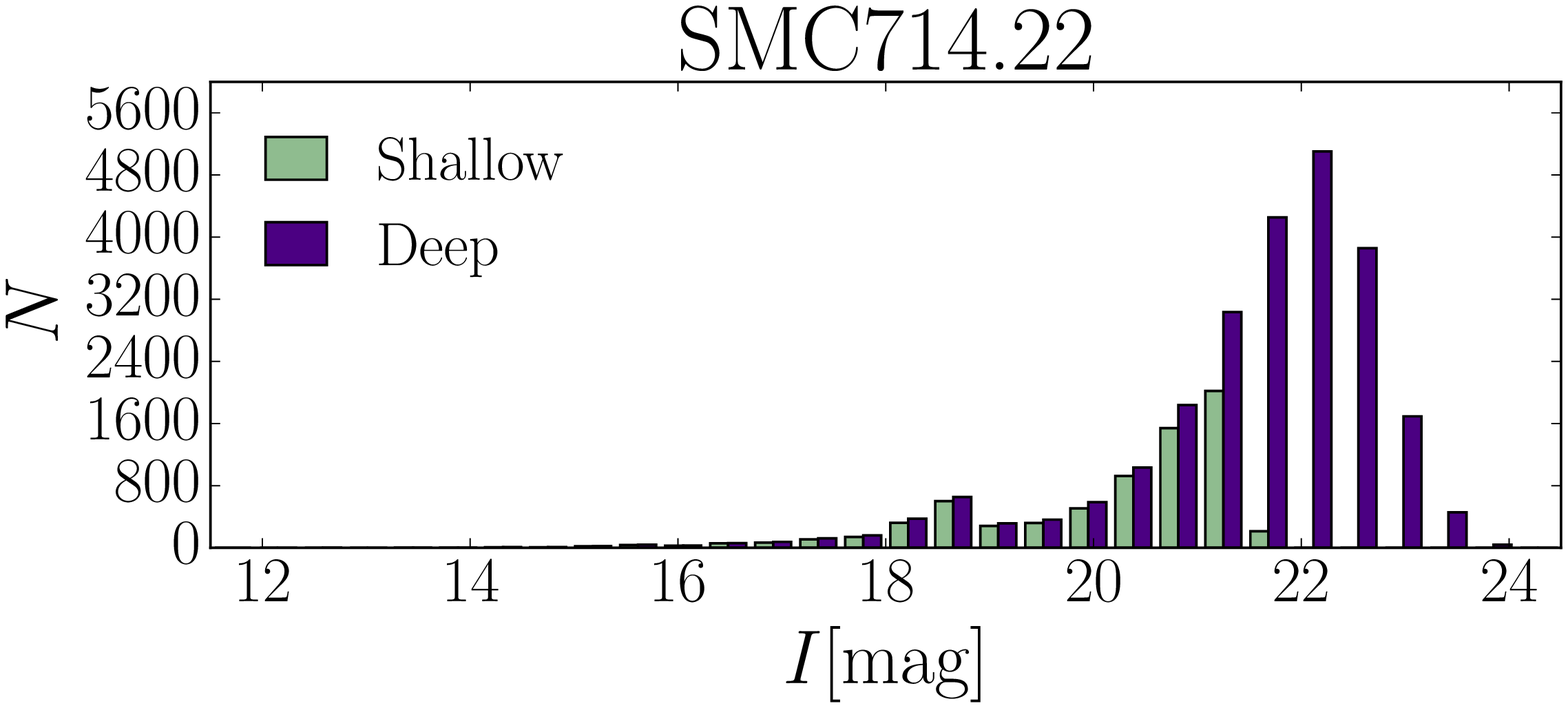}\hfill\includegraphics[width=6cm]{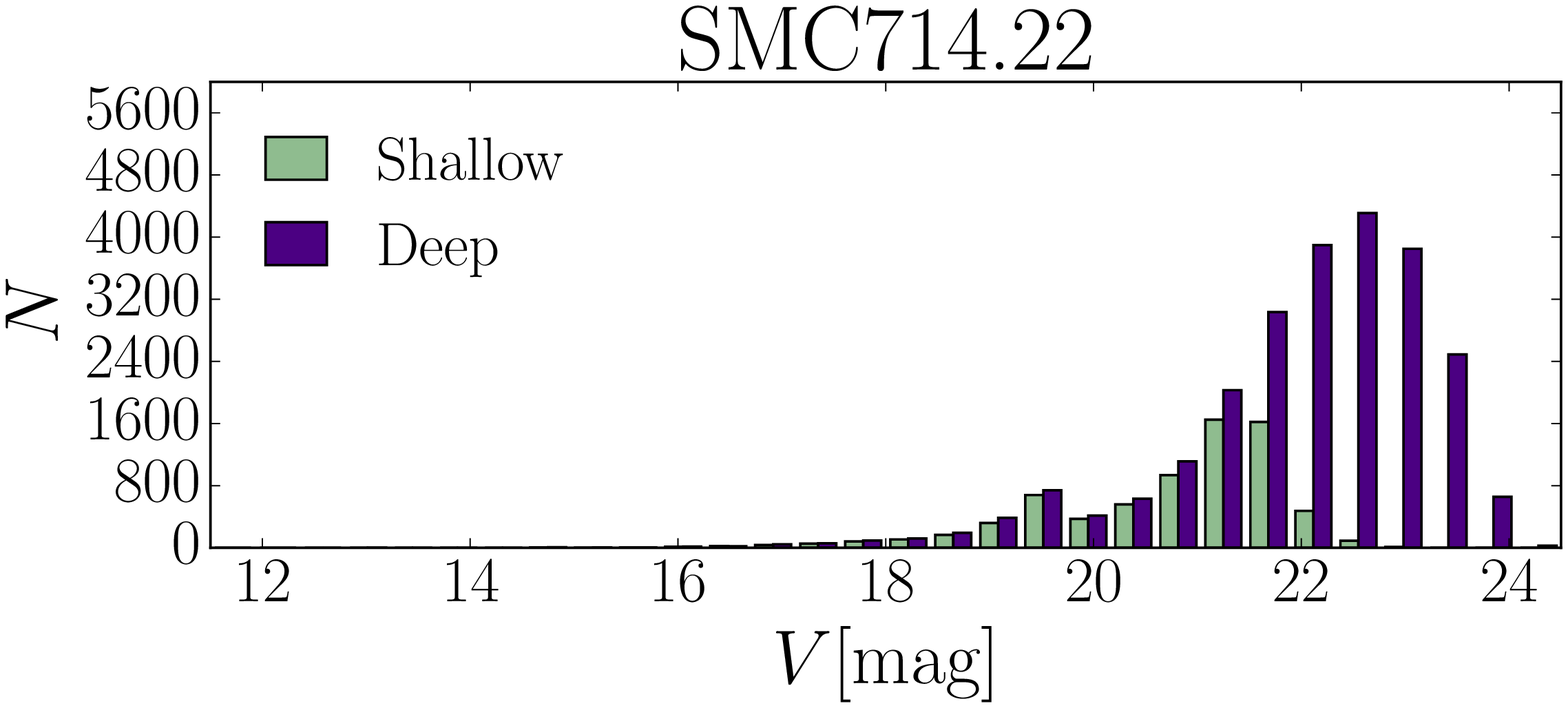}}
\centerline{\includegraphics[width=6cm]{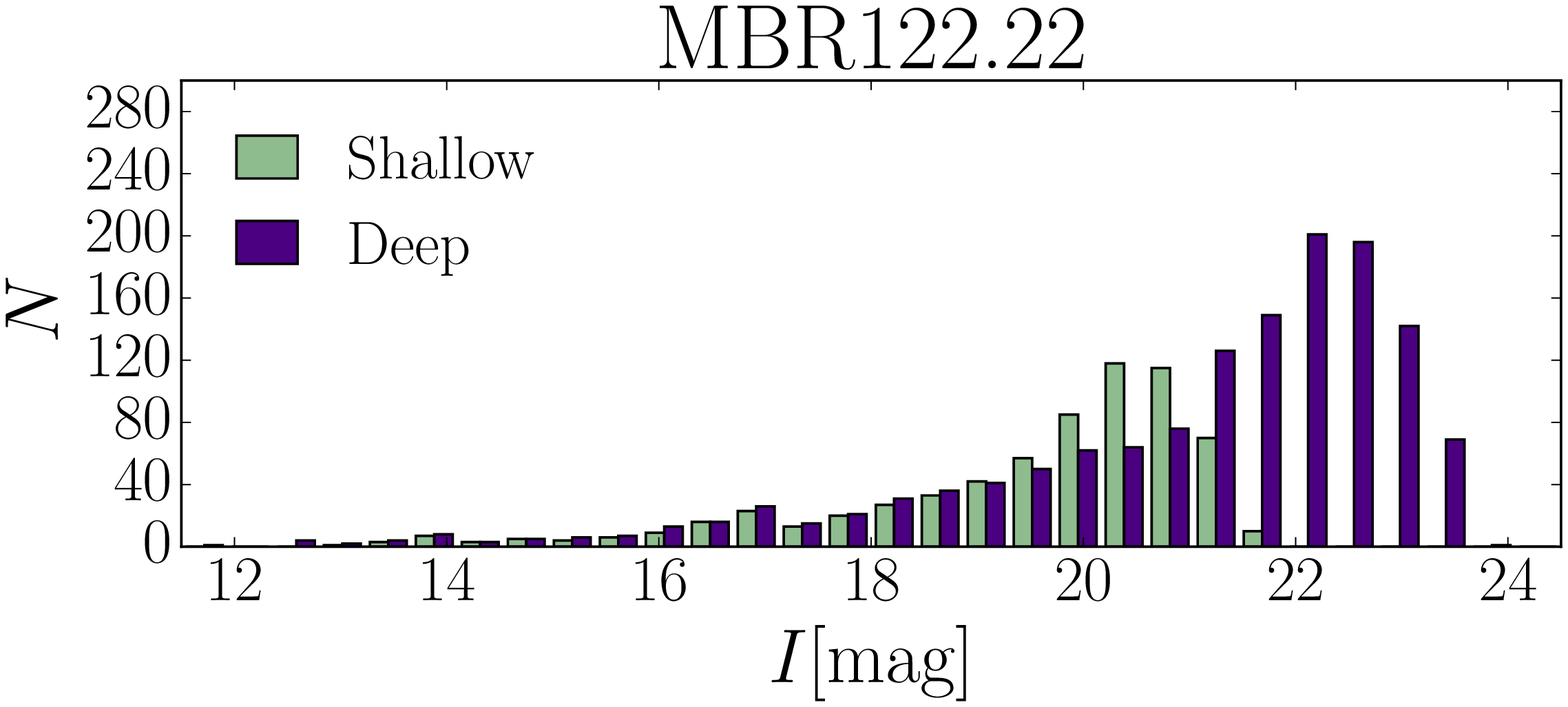}\hfill\includegraphics[width=6cm]{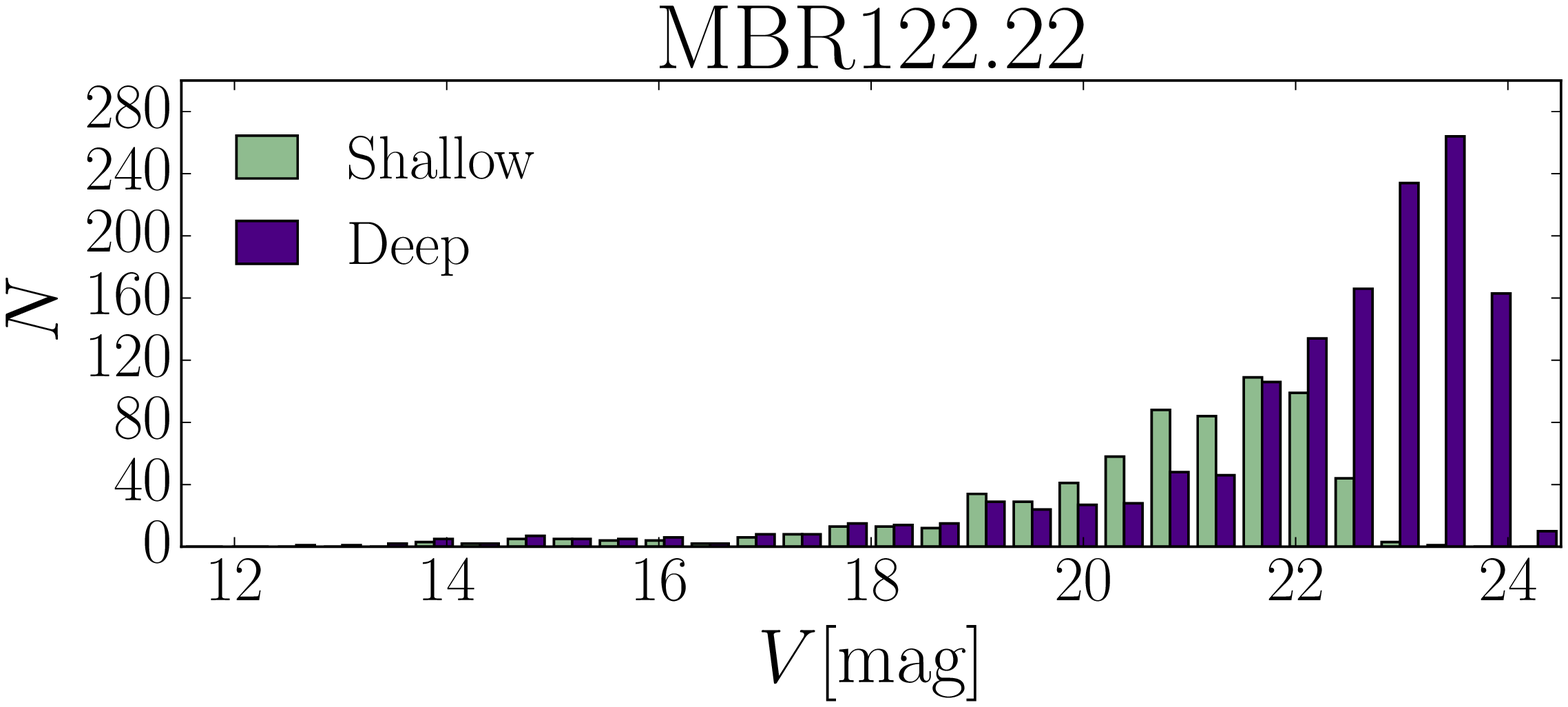}}
\vskip5pt
      \FigCap{Histograms of brightness ({\it left panels} are {\it I}-band
        and {\it right panels} are {\it V}-band) for two OGLE-IV subfields
        SMC714.22 ({\it top panels}) and MBR122.22 ({\it bottom panels}).}
\end{figure}
\begin{figure}[p]
\vglue-7mm
\centerline{\includegraphics[width=11cm, bb=0 10 180 200, clip=]{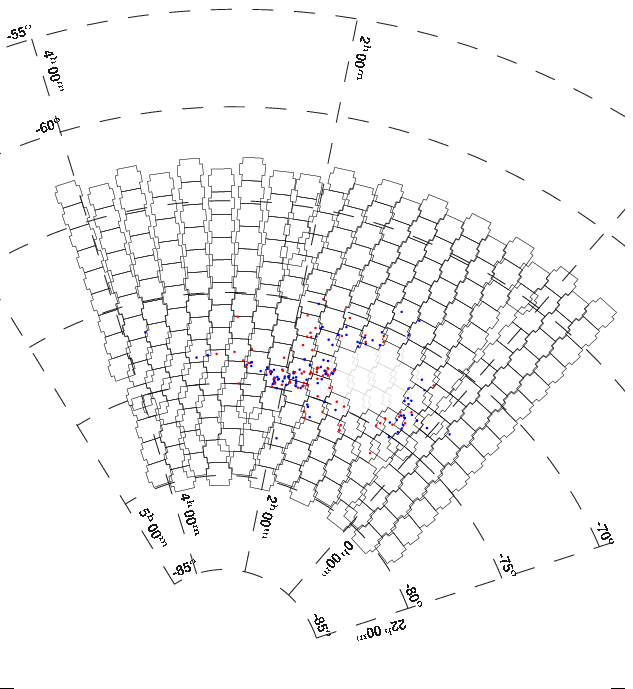}}
\FigCap{OGLE-IV fields in the SMC and MBR region. All outer black polygons
  were analyzed in this paper. Red and blue dots mark the location of newly
  discovered and the previously known star clusters, respectively (see
  Section~4).}
\vglue-.8cm
\centerline{\includegraphics[width=9cm, angle=270]{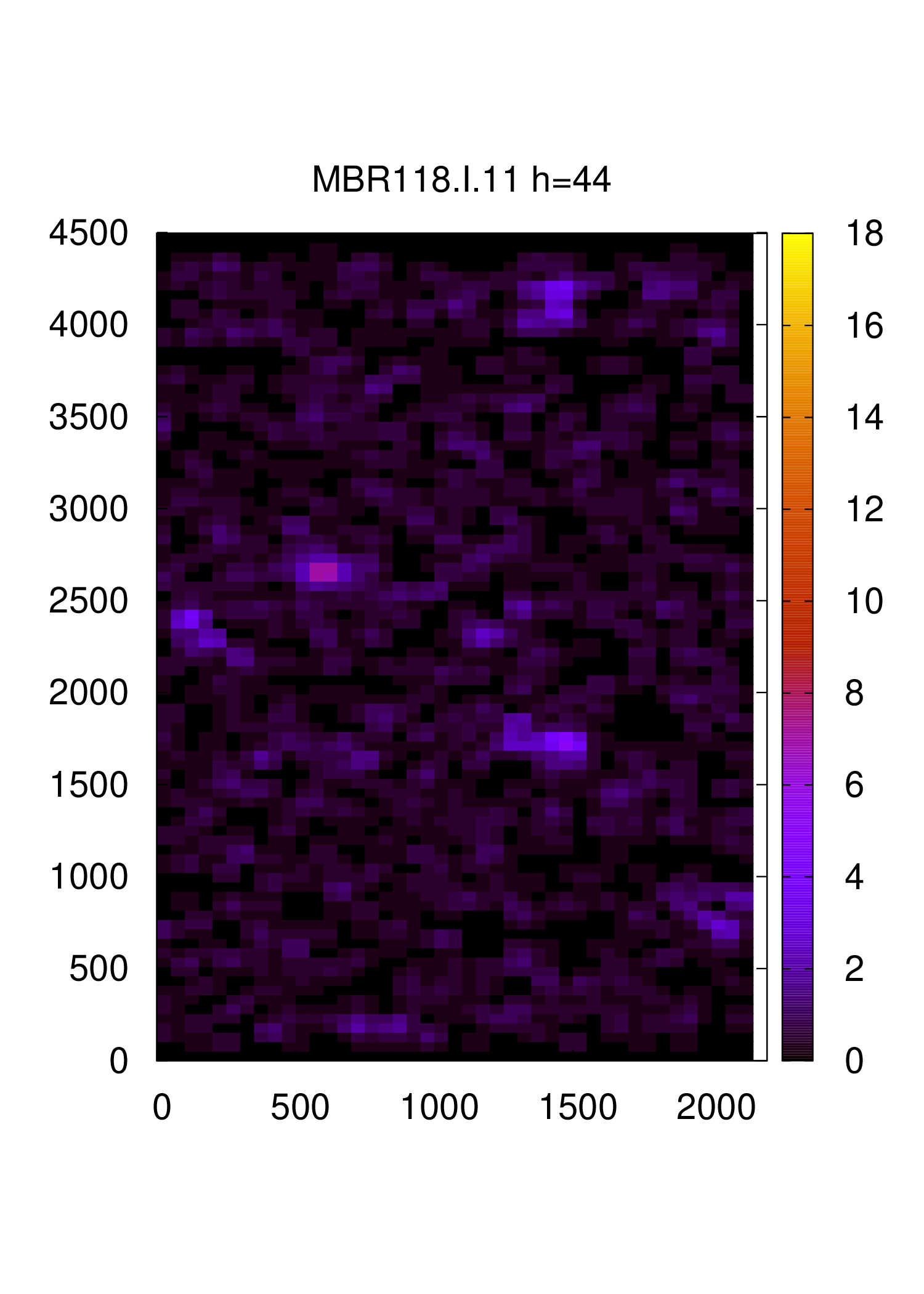}\hfill\includegraphics[width=9cm, angle=270]{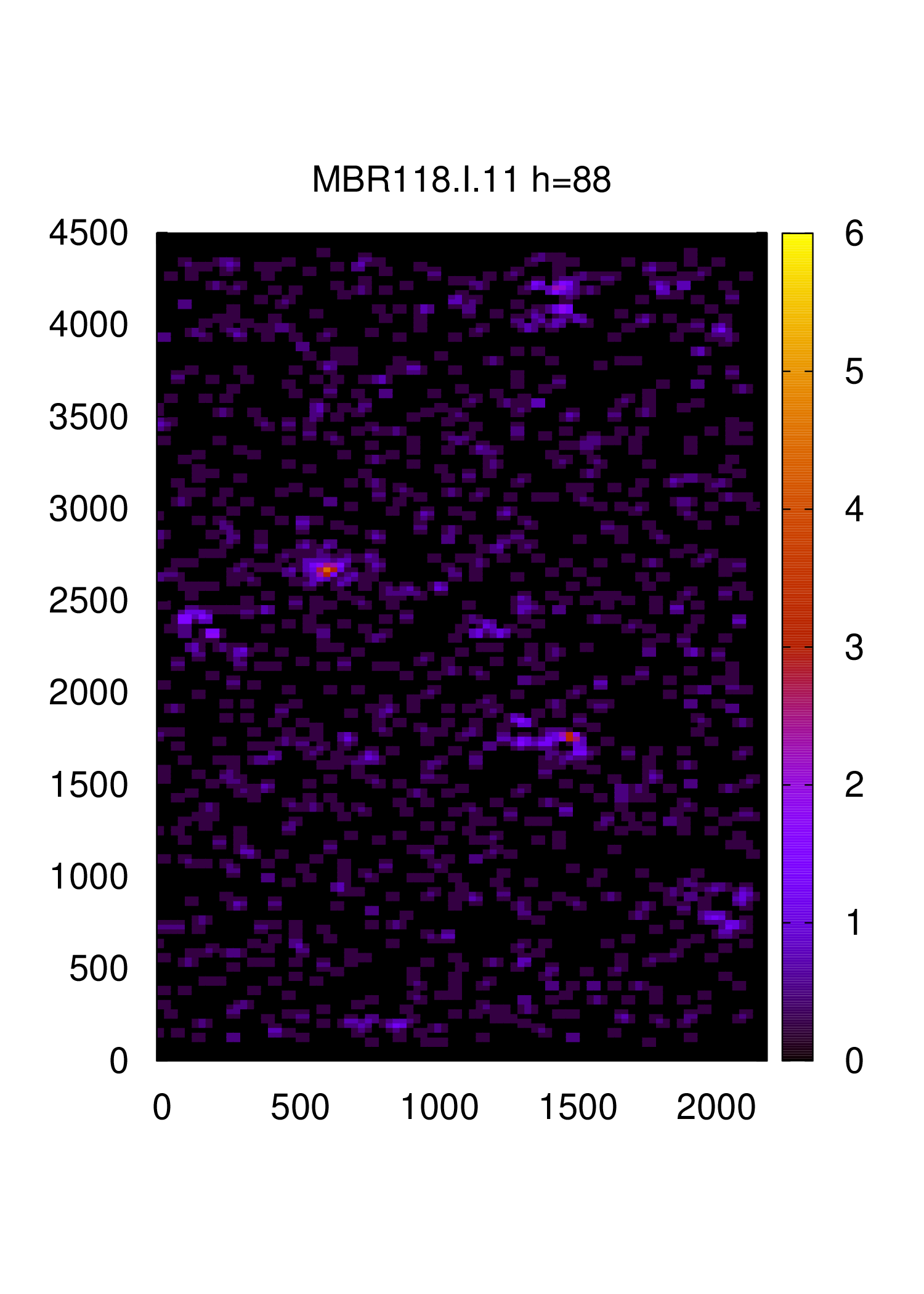}}
\vglue-.9cm
\FigCap{Stellar density maps of the MBR118.11 subfield for two
different cell sizes. The field contains new cluster: OGLE-MBR-CL-0033, 
which is located around (630,2680).}
\end{figure}

The star detection limit of the deep photometric maps of the SMC and MBR
reaches $I\approx23.5$~mag and $V\approx 24$~mag. The maps are complete to
about 22--23~mag in the {\it I}-band and 22.5--23.5 mag in the {\it
  V}-band, depending on the location and crowding of the field. These
limits are determined from the histograms of the mean magnitudes of the
stars by estimating the value where the numbers start to deviate from the
systematic growth (Fig.~1). All the details about observations, data
reductions and construction of the deep photometric maps can be found in
Paper~I.

\Section{Search for Clusters}
The method used in this paper is well established. The first automated
search of star clusters was performed by Zaritsky \etal (1997) and has been
used ever since. We used Zaritsky's method with small modifications which
are described in Paper~I (Section~3). Here, we present the analysis of the
fields located outside the central part of the SMC and the MBR (Fig.~2).

The examined area of 353 square degrees contains 252 OGLE-IV fields (each
field has 32 subfields what gives 8064 single subfields).  All analyzed
fields are shown in Fig.~2 as black polygons. The 10 gray polygons mark the
central SMC fields which have not been analyzed here. The list of all
analyzed SMC and MBR fields and their central coordinates are available on
the OGLE Web page\footnote{\it http://ogle.astrouw.edu.pl} together with other
supplementary information.

Exemplary density maps are presented in Figs.~3 and~4 for both MBR and SMC
subfields, respectively.  As in Paper I, we constructed a false-color
composition of images taken in the {\it I}- and {\it V}-bands (Fig.~5) and
plotted a photometric map of the region $400\times400$ pixels
($1\zdot\arcm7\times1\zdot\arcm7$) around each star cluster
candidate detected by our algorithm (see an example in Fig.~6).

Maps presented in Fig.~6 were made for the same object named
OGLE-MBR-CL-0033 which is presented in Fig.~5 and shown on the density map in Fig.~3.
\begin{figure}[p]
\vglue-1.1cm
\centerline{\includegraphics[width=9cm, angle=270]{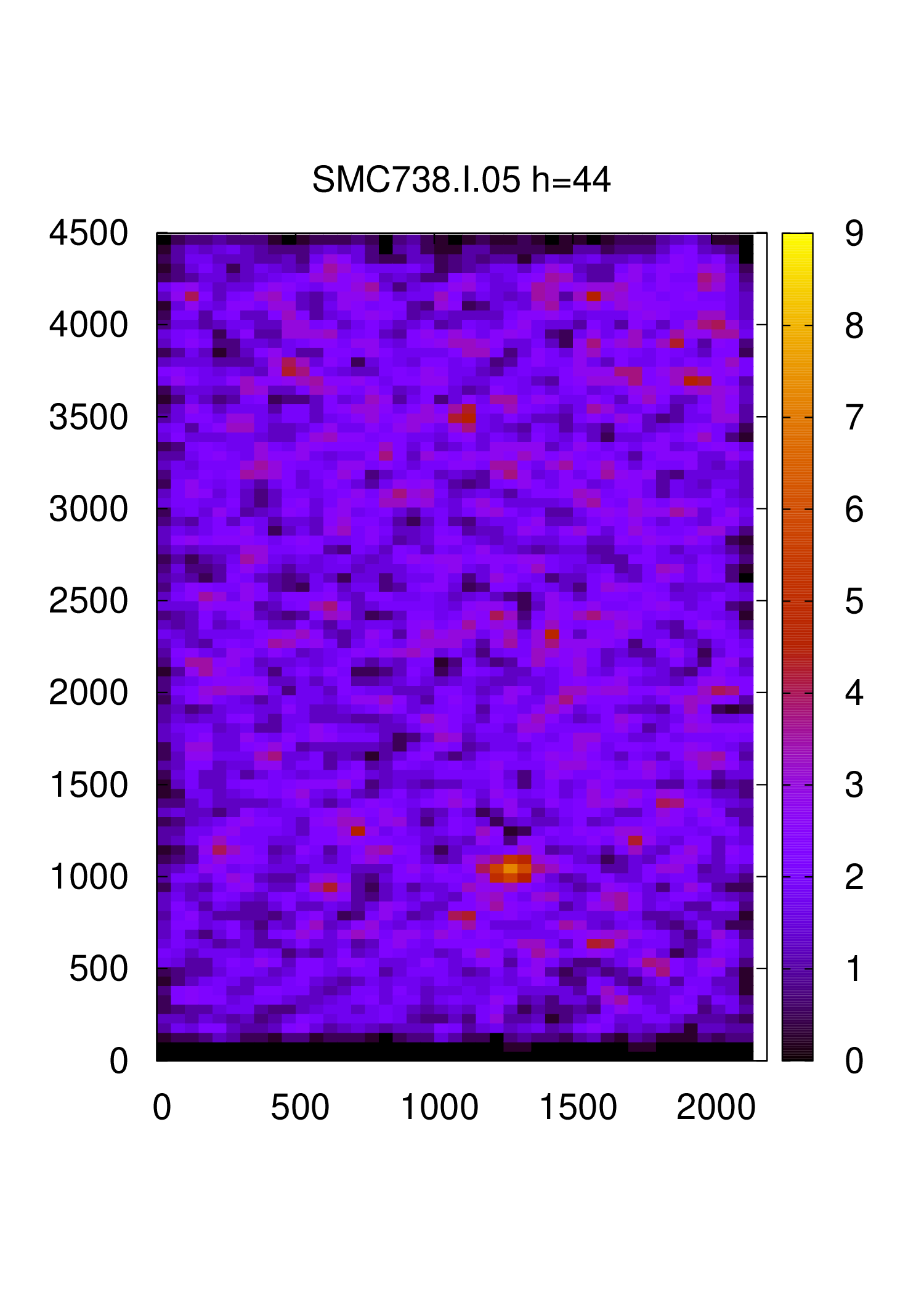}\hfill\includegraphics[width=9cm, angle=270]{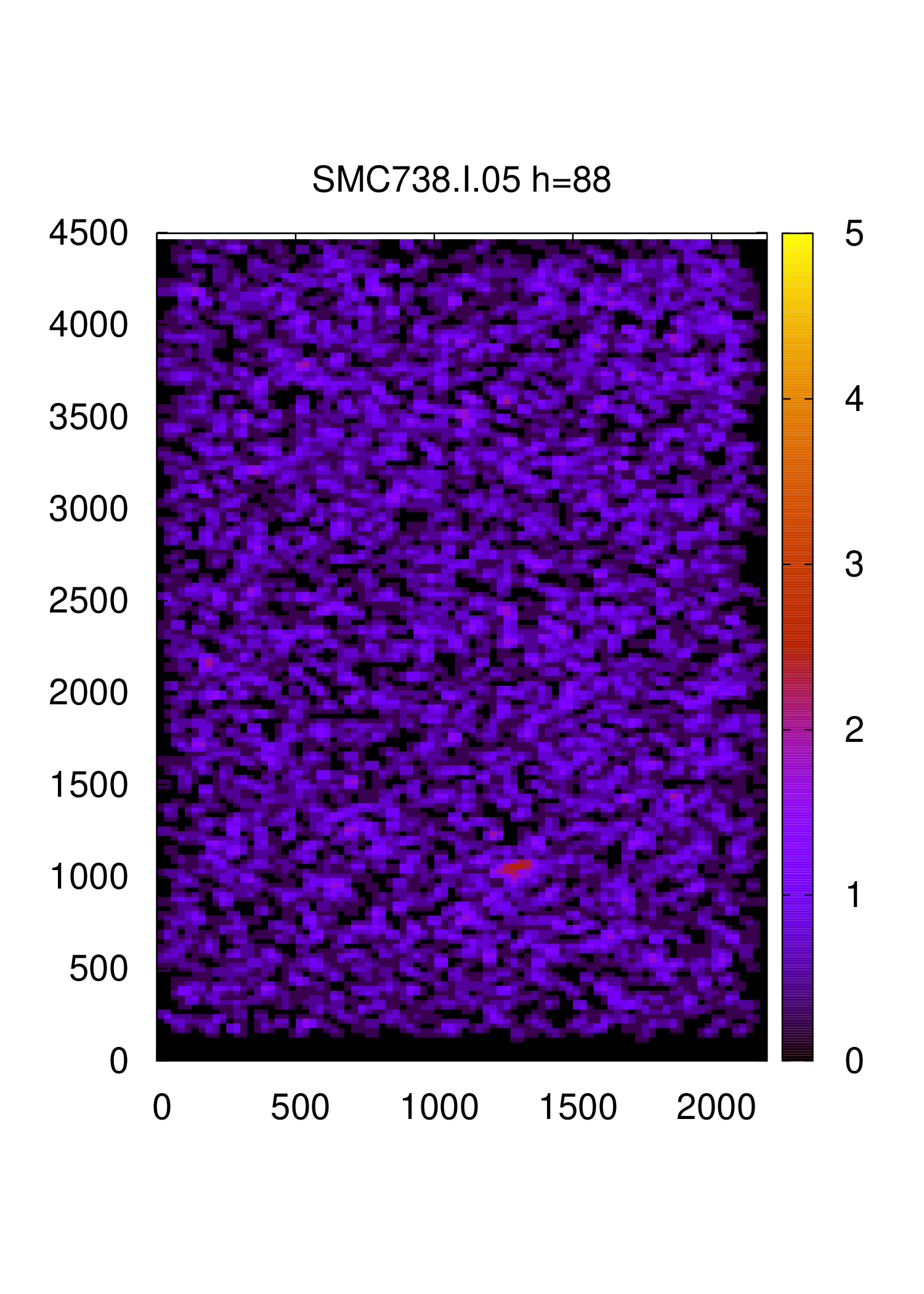}}
\vglue-1cm
\FigCap{Stellar density maps of a subfield SMC738.05 with object OGLE-SMC-CL-0242 
located around (1335,1090). This is a new candidate for star cluster.}
\vskip13mm
  \centerline{\includegraphics[width=12.3cm]{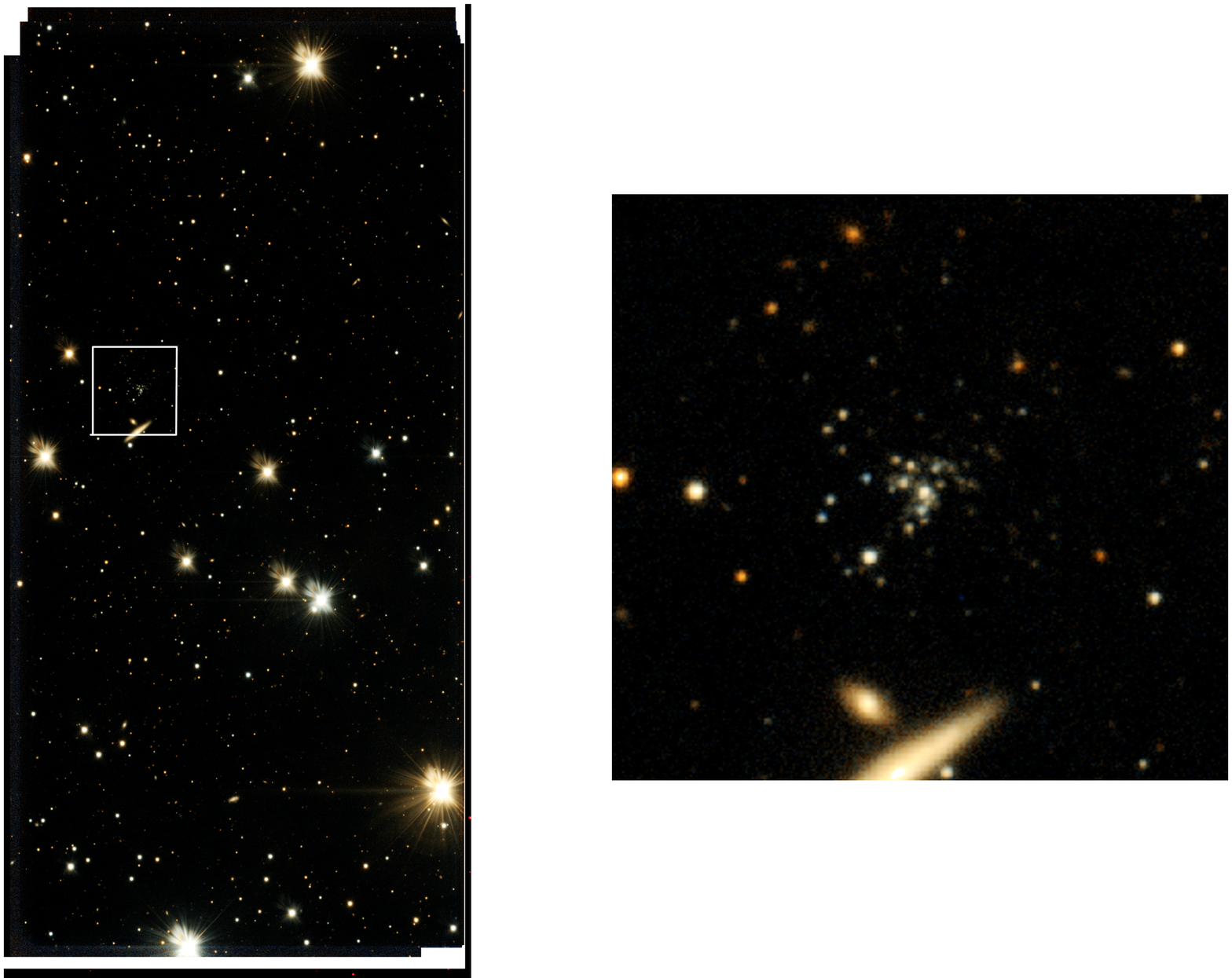}} 
\vskip13pt
  \FigCap{Color image of the subfield MBR118.11. The white square
    ($1\zdot\arcm7\times1\zdot\arcm7$) is enlarged in the {\it right panel},
    clearly showing the cluster OGLE-MBR-CL-0033 which is a new object.}
\vskip7mm
\end{figure}
\begin{figure}[htb]
\includegraphics[width=7.5cm]{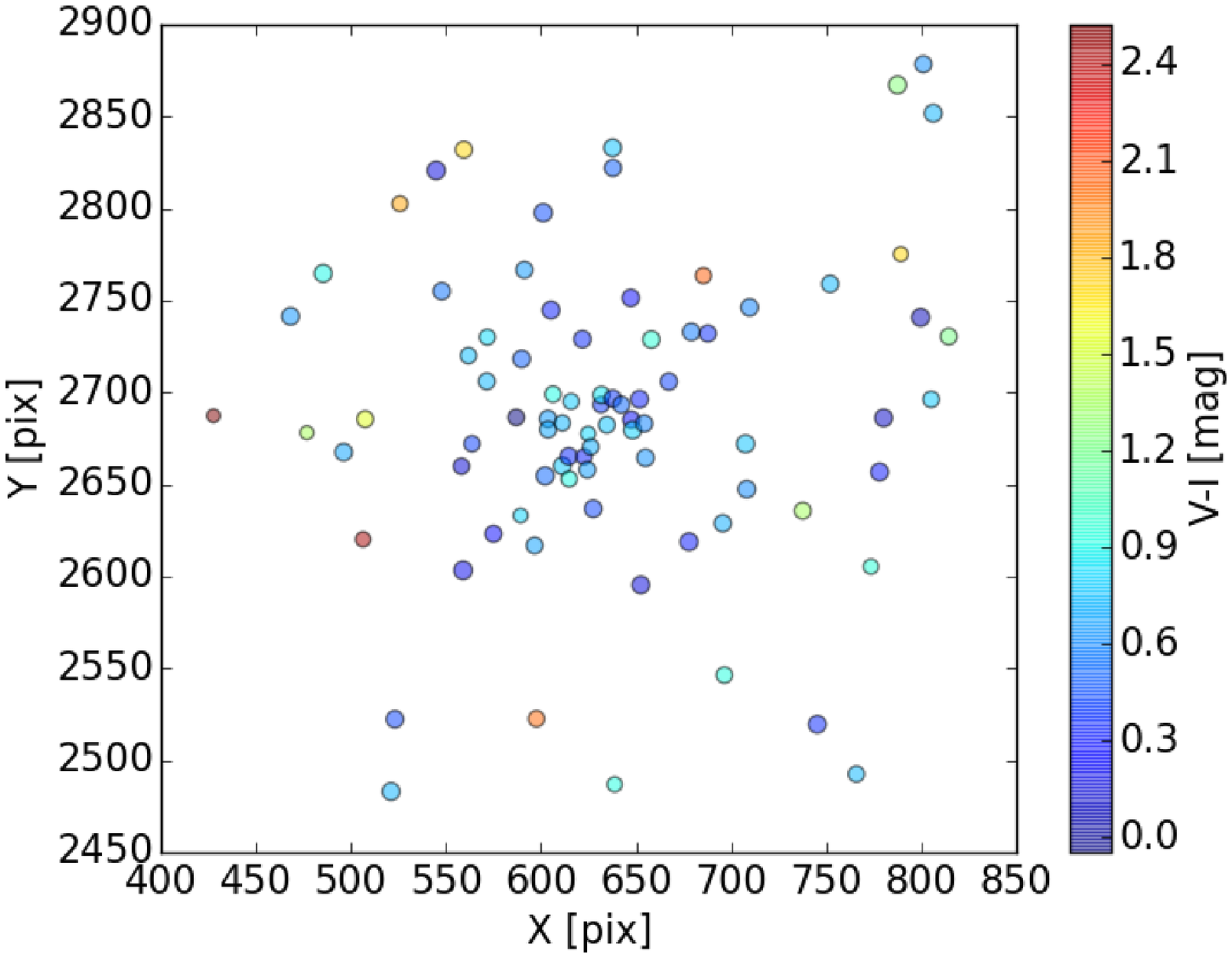}\hglue-7mm\includegraphics[width=7.5cm]{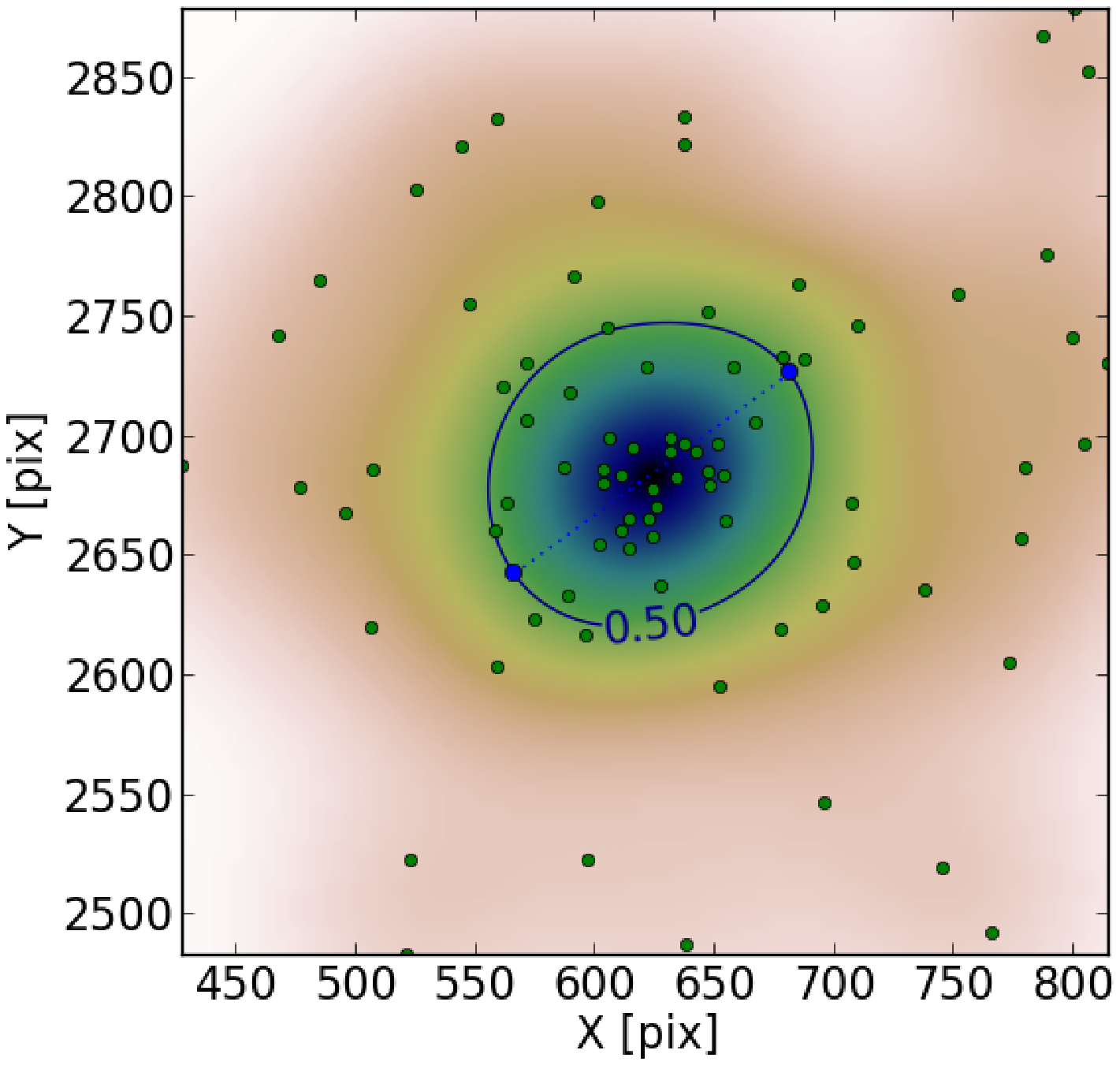}
\vskip11pt \FigCap{Deep photometric map of the cluster OGLE-MBR-CL-0033
  with the same size as in Fig.~5 -- $1\zdot\arcm7\times1\zdot\arcm7$. {\it
    Left panel} -- size of each point is proportional to the brightness of
  a star in the {\it I}-band, and the color represents the $V-I$ index of
  each star. {\it Right panel} -- deep photometric map with standard KDE
  distribution which was used to estimate the object's centroid.}
\end{figure}

For all regions found by our algorithm to be denser than the median
value for a given subfield we made a visual inspection, as described
in Paper~I.  For all objects which passed the visual inspection based
on six different images, the reliability index was assigned: 34\%
cluster candidates received the maximum value of 1, 28\% -- 0.9, 17\%
-- 0.8, 12\% -- 0.7 and 9\% -- 0.6. This index depends on the quantity
of images the object was identified on. The object received the
reliability index equal 1 if it was found on every image. The index
was reduced by 0.1 each time the object was not found on the image
from the inspected group. Object which was found only on two from six
images was rejected. All pictures (images in the {\it V}- and {\it I}-
bands, color images and a photometric map) of the accepted star
clusters are shown on the Web page (Section~4).

\vspace*{3pt} 
All the centroids were calculated in $XY$ coordinates of the
field and then converted to the equatorial coordinates. The estimation of
the approximate size of star cluster candidate was made using Kernel
Density Estimation (KDE) contour line at half maximum value (see an example
in Fig.~6. right panel). The calculations are the same as in
Paper~I. Table~1 presents the cluster parameters for the new objects and
Table~2 for the already known objects.

\vspace*{9pt}
\Section{The OGLE Collection of Star Clusters in the Outer Regions of the SMC and in the MBR}
\vspace*{5pt} 
We have found 198 star clusters in the outer regions of the SMC and in
the MBR. Among these, we have found 35 new star clusters in the 185
square degrees area of the MBR and 40 new star clusters in the 168
square degrees area of the outer regions of the SMC, based on
observations collected by the OGLE-IV survey. Their on-sky locations
are shown in Fig.~1 with red dots.  The remaining 123 objects were
identified in previously published catalogs -- 121 objects were listed
in the Bica catalog and two objects in Piatti (2017). They are marked
in Fig.~1 with blue dots.  As some extended objects cannot be
unambiguously classified, we have performed a cross-match of our
sample to both star clusters and associations from the Bica \etal
(2008) catalog. Some of their A-type objects (associations) were found
by our algorithm as clusters (the classification is shown in column~6
in Table~2). Almost all previously known objects located in the area
of the analyzed OGLE fields were detected by our algorithm, proving
its effectiveness and the completeness of the sample. There were eight
Bica objects (three star clusters and five clusters similar to
associations) which were not detected by our algorithm.  Those which
have Bica C-type were on the edge of a subfield, and those classified
as CA by Bica are not visible in our data. There was also one cluster
from Piatti (2017), Field16-02, which was found on the edge of our
subfield but rejected after visual inspection. All undetected objects
are listed in Table~3.

\vspace*{7pt} 
All discovered cluster candidates were numbered according to the
OGLE-IV naming scheme, which was presented in Paper~I. The name is
constructed as OGLE-MBR-CL-NNNN for the Magellanic Bridge and
OGLE-SMC-CL-NNNN for the Small Magellanic Cloud, where NNNN is an
object number. To make the numbering consistent with the OGLE-II
catalog (Pietrzyñski \etal 1998), we started it with 0239 for the
SMC. The MBR was not observed during previous OGLE phases so for the
MBR we started numbering with 0001.

\vspace*{7pt}
Table~1 presents the OGLE collection of new candidates for star
clusters. Column~1 contains the OGLE identification number, column~2 shows
the field name, in columns~3 and 4 we list the equatorial coordinates
(J2000) of the cluster center, column~5 contains the size of the cluster
(radius) in arcseconds and the last column contains the reliability index.

\vspace*{7pt} 
Table~2 presents the OGLE collection of star clusters, which
were already known. We cross-matched our detections with the Bica catalog
of star clusters as well as with the catalog of associations (both are part
of the Bica \etal 2008). Column~1 contains the OGLE identification number,
column~2 shows the field name. Columns 3 and 4 give our estimations of the
center equatorial coordinates (J2000).  Column 5 shows our estimation of
the size in arcseconds (radius), in column~6 we list the
cross-identification of extended objects. Column~7 contains the object type
from Bica \etal (2008): C -- ordinary cluster, CN -- cluster in nebula, CA
-- cluster similar to association, A -- ordinary association, AC --
association similar to cluster. Some of the objects have more than one
name or type because of problems with unambiguous cross-identification. Data
in Tables 1 and 2 show exactly the same parameters as in Paper~I to make
our collection self consistent.

\renewcommand{\arraystretch}{1.2}
\MakeTableSep{|c|c|c|c|c|c|}{12.5cm}{New star clusters}
{\hline
OGLE-IV name & OGLE-IV field & RA  & DEC & R$_{\rm KDE}$ [\arcs] & reliability \\
\hline
OGLE-MBR-CL-0001 & MBR100.32 & 1\uph46\upm02\zdot\ups68 & $-69\arcd42\arcm29\zdot\arcs9$ & 32 & 0.7 \\ 
OGLE-MBR-CL-0002 & MBR101.04 & 1\uph50\upm17\zdot\ups42 & $-71\arcd53\arcm28\zdot\arcs1$ & 49 & 0.7 \\ 
OGLE-MBR-CL-0003 & MBR101.05 & 1\uph48\upm33\zdot\ups32 & $-71\arcd38\arcm22\zdot\arcs0$ & 50 & 0.7 \\ 
OGLE-MBR-CL-0004 & MBR101.06 & 1\uph47\upm25\zdot\ups41 & $-71\arcd51\arcm11\zdot\arcs4$ & 48 & 0.7 \\ 
OGLE-MBR-CL-0005 & MBR101.15 & 1\uph46\upm01\zdot\ups10 & $-71\arcd19\arcm14\zdot\arcs9$ & 43 & 0.8 \\ 
OGLE-MBR-CL-0006 & MBR101.28 & 1\uph53\upm12\zdot\ups39 & $-70\arcd45\arcm09\zdot\arcs7$ & 29 & 0.9 \\ 
OGLE-MBR-CL-0007 & MBR102.15 & 1\uph44\upm09\zdot\ups39 & $-72\arcd31\arcm21\zdot\arcs2$ & 36 & 0.9 \\ 
OGLE-MBR-CL-0008 & MBR102.20 & 1\uph51\upm37\zdot\ups98 & $-72\arcd23\arcm22\zdot\arcs4$ & 34 & 0.7 \\ 
OGLE-MBR-CL-0009 & MBR103.10 & 1\uph52\upm44\zdot\ups33 & $-73\arcd55\arcm34\zdot\arcs7$ & 24 & 1 \\ 
OGLE-MBR-CL-0010 & MBR103.13 & 1\uph47\upm29\zdot\ups34 & $-73\arcd53\arcm39\zdot\arcs5$ & 37 & 0.9 \\ 
OGLE-MBR-CL-0011 & MBR103.13 & 1\uph45\upm54\zdot\ups59 & $-73\arcd46\arcm50\zdot\arcs0$ & 44 & 0.8 \\ 
OGLE-MBR-CL-0012 & MBR103.16 & 1\uph41\upm06\zdot\ups00 & $-73\arcd49\arcm00\zdot\arcs9$ & 22 & 0.9 \\ 
OGLE-MBR-CL-0013 & MBR103.25 & 1\uph40\upm39\zdot\ups05 & $-73\arcd43\arcm05\zdot\arcs8$ & 27 & 1 \\ 
OGLE-MBR-CL-0014 & MBR104.21 & 1\uph47\upm29\zdot\ups66 & $-74\arcd42\arcm54\zdot\arcs8$ & 33 & 1 \\ 
OGLE-MBR-CL-0015 & MBR104.26 & 1\uph54\upm37\zdot\ups56 & $-74\arcd32\arcm42\zdot\arcs9$ & 36 & 0.9 \\ 
OGLE-MBR-CL-0016 & MBR104.26 & 1\uph54\upm46\zdot\ups63 & $-74\arcd29\arcm04\zdot\arcs3$ & 27 & 0.9 \\ 
OGLE-MBR-CL-0017 & MBR104.31 & 1\uph43\upm28\zdot\ups27 & $-74\arcd31\arcm45\zdot\arcs7$ & 42 & 0.8 \\ 
OGLE-MBR-CL-0018 & MBR104.31 & 1\uph43\upm50\zdot\ups47 & $-74\arcd22\arcm12\zdot\arcs1$ & 38 & 0.8 \\ 
OGLE-MBR-CL-0019 & MBR104.32 & 1\uph40\upm29\zdot\ups43 & $-74\arcd26\arcm05\zdot\arcs3$ & 31 & 1 \\ 
OGLE-MBR-CL-0020 & MBR105.25 & 1\uph35\upm04\zdot\ups87 & $-76\arcd11\arcm19\zdot\arcs6$ & 21 & 0.9 \\ 
OGLE-MBR-CL-0021 & MBR108.13 & 2\uph03\upm13\zdot\ups01 & $-73\arcd16\arcm49\zdot\arcs3$ & 25 & 0.9 \\ 
OGLE-MBR-CL-0022 & MBR109.03 & 2\uph08\upm40\zdot\ups16 & $-74\arcd57\arcm28\zdot\arcs7$ & 45 & 0.7 \\ 
OGLE-MBR-CL-0023 & MBR109.04 & 2\uph06\upm38\zdot\ups84 & $-74\arcd45\arcm17\zdot\arcs0$ & 21 & 0.6 \\ 
OGLE-MBR-CL-0024 & MBR109.06 & 2\uph00\upm45\zdot\ups78 & $-74\arcd45\arcm07\zdot\arcs6$ & 35 & 0.8 \\ 
OGLE-MBR-CL-0025 & MBR109.11 & 2\uph07\upm38\zdot\ups63 & $-74\arcd29\arcm42\zdot\arcs8$ & 32 & 1 \\ 
OGLE-MBR-CL-0026 & MBR109.18 & 2\uph12\upm11\zdot\ups91 & $-74\arcd10\arcm23\zdot\arcs2$ & 28 & 0.6 \\ 
OGLE-MBR-CL-0027 & MBR109.31 & 2\uph02\upm12\zdot\ups63 & $-73\arcd59\arcm32\zdot\arcs5$ & 31 & 0.9 \\ 
OGLE-MBR-CL-0028 & MBR112.01 & 2\uph28\upm02\zdot\ups11 & $-73\arcd02\arcm53\zdot\arcs4$ & 23 & 0.8 \\ 
OGLE-MBR-CL-0029 & MBR113.08 & 2\uph31\upm30\zdot\ups69 & $-73\arcd59\arcm26\zdot\arcs9$ & 21 & 1 \\ 
OGLE-MBR-CL-0030 & MBR113.09 & 2\uph29\upm53\zdot\ups33 & $-73\arcd49\arcm10\zdot\arcs8$ & 25 & 1 \\ 
OGLE-MBR-CL-0031 & MBR117.27 & 2\uph39\upm28\zdot\ups54 & $-71\arcd19\arcm02\zdot\arcs7$ & 36 & 0.6 \\ 
OGLE-MBR-CL-0032 & MBR118.06 & 2\uph34\upm08\zdot\ups09 & $-73\arcd42\arcm14\zdot\arcs6$ & 19 & 0.8 \\ 
OGLE-MBR-CL-0033 & MBR118.11 & 2\uph41\upm03\zdot\ups58 & $-73\arcd15\arcm12\zdot\arcs4$ & 17 & 1 \\ 
OGLE-MBR-CL-0034 & MBR119.06 & 2\uph36\upm38\zdot\ups95 & $-74\arcd58\arcm21\zdot\arcs0$ & 24 & 0.6 \\ 
OGLE-MBR-CL-0035 & MBR123.29 & 2\uph54\upm42\zdot\ups98 & $-73\arcd22\arcm17\zdot\arcs4$ & 28 & 0.9 * \\ 
OGLE-SMC-CL-0239 & SMC738.01 & 1\uph40\upm38\zdot\ups72 & $-73\arcd43\arcm03\zdot\arcs6$ & 28 & 1 \\ 
OGLE-SMC-CL-0240 & SMC738.02 & 1\uph37\upm02\zdot\ups76 & $-73\arcd28\arcm18\zdot\arcs6$ & 27 & 0.8 \\ 
\hline
\noalign{\vskip5pt}
\multicolumn{6}{p{12.5cm}}{*Object OGLE-MBR-CL-0035 was detected on the edge of the known association ICA65ne.}
}
\setcounter{table}{0}
\MakeTableSep{|c|c|c|c|c|c|}{12.5cm}{New star clusters}
{\hline
OGLE-IV name & OGLE-IV field & RA  & DEC & R$_{\rm KDE}$ [\arcs] & reliability \\
\hline
OGLE-SMC-CL-0241 & SMC738.03 & 1\uph36\upm24\zdot\ups26 & $-73\arcd36\arcm30\zdot\arcs8$ & 37 & 0.9 \\ 
OGLE-SMC-CL-0242 & SMC738.05 & 1\uph31\upm13\zdot\ups45 & $-73\arcd41\arcm35\zdot\arcs3$ & 26 & 1 \\ 
OGLE-SMC-CL-0243 & SMC738.07 & 1\uph27\upm35\zdot\ups51 & $-73\arcd33\arcm58\zdot\arcs2$ & 44 & 0.9 \\ 
OGLE-SMC-CL-0244 & SMC738.07 & 1\uph27\upm41\zdot\ups17 & $-73\arcd35\arcm49\zdot\arcs0$ & 48 & 0.6 \\ 
OGLE-SMC-CL-0245 & SMC738.10 & 1\uph37\upm15\zdot\ups96 & $-73\arcd22\arcm53\zdot\arcs8$ & 37 & 0.9 \\ 
OGLE-SMC-CL-0246 & SMC738.11 & 1\uph36\upm19\zdot\ups48 & $-73\arcd21\arcm49\zdot\arcs5$ & 44 & 0.8 \\ 
OGLE-SMC-CL-0247 & SMC738.14 & 1\uph28\upm57\zdot\ups40 & $-73\arcd14\arcm55\zdot\arcs7$ & 38 & 1 \\ 
OGLE-SMC-CL-0248 & SMC738.14 & 1\uph29\upm49\zdot\ups16 & $-73\arcd25\arcm48\zdot\arcs7$ & 45 & 0.8 \\ 
OGLE-SMC-CL-0249 & SMC738.14 & 1\uph29\upm36\zdot\ups38 & $-73\arcd13\arcm39\zdot\arcs7$ & 38 & 0.8 \\ 
OGLE-SMC-CL-0250 & SMC738.16 & 1\uph24\upm33\zdot\ups75 & $-73\arcd14\arcm21\zdot\arcs2$ & 44 & 0.9 \\ 
OGLE-SMC-CL-0251 & SMC738.16 & 1\uph24\upm38\zdot\ups61 & $-73\arcd12\arcm57\zdot\arcs6$ & 47 & 0.9 \\ 
OGLE-SMC-CL-0252 & SMC739.08 & 1\uph40\upm30\zdot\ups50 & $-74\arcd26\arcm04\zdot\arcs9$ & 28 & 1 \\ 
OGLE-SMC-CL-0253 & SMC739.25 & 1\uph23\upm11\zdot\ups37 & $-74\arcd12\arcm06\zdot\arcs0$ & 37 & 0.7 \\ 
OGLE-SMC-CL-0254 & SMC739.25 & 1\uph22\upm52\zdot\ups99 & $-74\arcd11\arcm28\zdot\arcs5$ & 50 & 0.7 \\ 
OGLE-SMC-CL-0255 & SMC739.28 & 1\uph34\upm30\zdot\ups37 & $-73\arcd56\arcm59\zdot\arcs9$ & 31 & 1 \\ 
OGLE-SMC-CL-0256 & SMC740.09 & 1\uph36\upm41\zdot\ups78 & $-75\arcd51\arcm15\zdot\arcs0$ & 24 & 1 \\ 
OGLE-SMC-CL-0257 & SMC740.24 & 1\uph21\upm30\zdot\ups56 & $-75\arcd33\arcm15\zdot\arcs2$ & 39 & 1 \\ 
OGLE-SMC-CL-0258 & SMC734.24 & 1\uph06\upm44\zdot\ups77 & $-74\arcd49\arcm58\zdot\arcs5$ & 32 & 1 \\ 
OGLE-SMC-CL-0259 & SMC735.24 & 1\uph01\upm50\zdot\ups99 & $-76\arcd06\arcm38\zdot\arcs4$ & 27 & 0.8 \\ 
OGLE-SMC-CL-0260 & SMC735.31 & 1\uph02\upm36\zdot\ups01 & $-75\arcd49\arcm22\zdot\arcs7$ & 34 & 0.6 \\ 
OGLE-SMC-CL-0261 & SMC728.01 & 1\uph01\upm51\zdot\ups09 & $-76\arcd06\arcm35\zdot\arcs2$ & 23 & 0.9 \\ 
OGLE-SMC-CL-0262 & SMC730.10 & 1\uph26\upm31\zdot\ups46 & $-70\arcd16\arcm09\zdot\arcs4$ & 23 & 0.9 \\ 
OGLE-SMC-CL-0263 & SMC722.14 & 0\uph28\upm02\zdot\ups60 & $-76\arcd21\arcm12\zdot\arcs3$ & 19 & 1 \\ 
OGLE-SMC-CL-0264 & SMC706.16 & 0\uph18\upm22\zdot\ups14 & $-71\arcd27\arcm02\zdot\arcs2$ & 18 & 1 \\ 
OGLE-SMC-CL-0265 & SMC708.08 & 0\uph25\upm17\zdot\ups91 & $-73\arcd52\arcm10\zdot\arcs6$ & 39 & 1 \\ 
OGLE-SMC-CL-0266 & SMC708.11 & 0\uph18\upm56\zdot\ups07 & $-73\arcd57\arcm37\zdot\arcs4$ & 52 & 0.7 \\ 
OGLE-SMC-CL-0267 & SMC708.17 & 0\uph26\upm32\zdot\ups81 & $-73\arcd38\arcm07\zdot\arcs2$ & 53 & 0.6 \\ 
OGLE-SMC-CL-0268 & SMC708.28 & 0\uph19\upm48\zdot\ups95 & $-73\arcd18\arcm05\zdot\arcs7$ & 34 & 1 \\ 
OGLE-SMC-CL-0269 & SMC714.01 & 0\uph34\upm48\zdot\ups90 & $-74\arcd42\arcm18\zdot\arcs5$ & 52 & 1 \\ 
OGLE-SMC-CL-0270 & SMC714.13 & 0\uph26\upm04\zdot\ups88 & $-74\arcd24\arcm59\zdot\arcs5$ & 50 & 0.9 \\ 
OGLE-SMC-CL-0271 & SMC714.19 & 0\uph33\upm27\zdot\ups41 & $-74\arcd21\arcm42\zdot\arcs3$ & 36 & 1 \\ 
OGLE-SMC-CL-0272 & SMC721.23 & 0\uph36\upm31\zdot\ups49 & $-74\arcd55\arcm46\zdot\arcs7$ & 39 & 1 \\ 
OGLE-SMC-CL-0273 & SMC721.32 & 0\uph33\upm27\zdot\ups97 & $-74\arcd21\arcm45\zdot\arcs5$ & 41 & 0.9 \\ 
OGLE-SMC-CL-0274 & SMC724.02 & 1\uph11\upm23\zdot\ups08 & $-71\arcd10\arcm00\zdot\arcs4$ & 47 & 0.9 \\ 
OGLE-SMC-CL-0275 & SMC724.09 & 1\uph13\upm18\zdot\ups32 & $-70\arcd50\arcm45\zdot\arcs7$ & 40 & 1 \\ 
OGLE-SMC-CL-0276 & SMC724.16 & 1\uph00\upm46\zdot\ups51 & $-70\arcd47\arcm12\zdot\arcs8$ & 40 & 1 \\ 
OGLE-SMC-CL-0277 & SMC724.24 & 1\uph02\upm20\zdot\ups76 & $-70\arcd29\arcm08\zdot\arcs4$ & 37 & 0.8 \\ 
OGLE-SMC-CL-0278 & SMC731.09 & 1\uph27\upm30\zdot\ups34 & $-71\arcd19\arcm26\zdot\arcs5$ & 36 & 0.9 \\ 
\hline}

\begin{landscape}
\renewcommand{\arraystretch}{1.3}
\renewcommand{\TableFont}{\scriptsize}
\MakeTableSepp{|c|c|c|c|c|c|c|}{12.5cm}{Already known star clusters}
{\hline
OGLE-IV name & OGLE-IV field & RA & DEC & $R_{\rm KDE}$ [\arcs] & name & cluster type \\ 
\hline
OGLE-MBR-CL-0036 & MBR100.23 &  1\uph48\upm01\zdot\ups75 & $-70\arcd00\arcm13\zdot\arcs1$ & 20 & BS196 & C \\ 
OGLE-MBR-CL-0037 & MBR101.16 &  1\uph42\upm29\zdot\ups05 & $-71\arcd16\arcm52\zdot\arcs8$ & 17 & HW85 & C \\ 
OGLE-MBR-CL-0038 & MBR102.05 &  1\uph47\upm56\zdot\ups36 & $-73\arcd07\arcm38\zdot\arcs7$ & 35 & BS198 & CA \\ 
OGLE-MBR-CL-0039 & MBR103.01 &  1\uph56\upm44\zdot\ups64 & $-74\arcd13\arcm09\zdot\arcs9$ & 23 & NGC796,L115,WG9,ESO30SC6 & C \\ 
OGLE-MBR-CL-0040 & MBR103.02 &  1\uph52\upm57\zdot\ups31 & $-74\arcd14\arcm56\zdot\arcs7$ & 29 & BS207 & C \\ 
OGLE-MBR-CL-0041 & MBR103.03 &  1\uph50\upm20\zdot\ups50 & $-74\arcd21\arcm10\zdot\arcs3$ & 28 & L114,WG4,ESO30SC5 & C \\ 
OGLE-MBR-CL-0042 & MBR103.03 &  1\uph50\upm55\zdot\ups38 & $-74\arcd10\arcm43\zdot\arcs3$ & 38 & WG5se & CA \\ 
OGLE-MBR-CL-0043 & MBR103.07 &  1\uph42\upm23\zdot\ups53 & $-74\arcd10\arcm24\zdot\arcs7$ & 42 & HW86 & C \\ 
OGLE-MBR-CL-0044 & MBR103.10 &  1\uph53\upm48\zdot\ups21 & $-73\arcd56\arcm09\zdot\arcs3$ & 25 & BS212 & C \\ 
OGLE-MBR-CL-0045 & MBR103.10 &  1\uph53\upm34\zdot\ups18 & $-74\arcd00\arcm26\zdot\arcs7$ & 38 & BS210 & A \\ 
OGLE-MBR-CL-0046 & MBR103.10 &  1\uph53\upm12\zdot\ups57 & $-73\arcd58\arcm39\zdot\arcs6$ & 45 & WG6 & C \\ 
OGLE-MBR-CL-0047 & MBR103.21 &  1\uph49\upm30\zdot\ups93 & $-73\arcd43\arcm57\zdot\arcs0$ & 52 & L113,ESO30SC4 & C \\ 
OGLE-MBR-CL-0048 & MBR103.29 &  1\uph48\upm01\zdot\ups05 & $-73\arcd07\arcm55\zdot\arcs7$ & 23 & BS198 & CA \\ 
OGLE-MBR-CL-0049 & MBR103.32 &  1\uph42\upm53\zdot\ups16 & $-73\arcd20\arcm13\zdot\arcs6$ & 21 & WG1 & C \\ 
OGLE-MBR-CL-0050 & MBR104.17 &  1\uph57\upm16\zdot\ups53 & $-74\arcd42\arcm32\zdot\arcs0$ & 31 & BS218 & A \\ 
OGLE-MBR-CL-0051 & MBR104.22 &  1\uph45\upm14\zdot\ups28 & $-74\arcd41\arcm23\zdot\arcs3$ & 30 & WG2/BS195 & CA/A \\ 
OGLE-MBR-CL-0052 & MBR104.28 &  1\uph49\upm43\zdot\ups75 & $-74\arcd36\arcm55\zdot\arcs3$ & 26 & WG3 & CA \\ 
OGLE-MBR-CL-0053 & MBR104.28 &  1\uph49\upm25\zdot\ups56 & $-74\arcd39\arcm11\zdot\arcs5$ & 26 & BSBD3/BBDS2 & CN/AN \\ 
OGLE-MBR-CL-0054 & MBR104.28 &  1\uph49\upm52\zdot\ups27 & $-74\arcd28\arcm49\zdot\arcs0$ & 45 & BS202 & A \\ 
OGLE-MBR-CL-0055 & MBR104.28 &  1\uph50\upm18\zdot\ups00 & $-74\arcd21\arcm34\zdot\arcs3$ & 20 & L114,WG4,ESO30SC5 & C \\ 
OGLE-MBR-CL-0056 & MBR104.31 &  1\uph43\upm50\zdot\ups16 & $-74\arcd34\arcm16\zdot\arcs7$ & 41 & BS192 & CA \\ 
OGLE-MBR-CL-0057 & MBR104.31 &  1\uph43\upm53\zdot\ups64 & $-74\arcd32\arcm25\zdot\arcs2$ & 36 & BS193 & C \\ 
OGLE-MBR-CL-0058 & MBR109.03 &  2\uph08\upm19\zdot\ups40 & $-74\arcd48\arcm11\zdot\arcs2$ & 35 & WG16 & AC \\ 
OGLE-MBR-CL-0059 & MBR109.03 &  2\uph07\upm44\zdot\ups30 & $-74\arcd45\arcm44\zdot\arcs4$ & 41 & BS228 & AC \\ 
OGLE-MBR-CL-0060 & MBR109.04 &  2\uph06\upm50\zdot\ups82 & $-74\arcd41\arcm31\zdot\arcs4$ & 32 & ICA11 & A \\ 
\hline}
\setcounter{table}{1}
\MakeTableSepp{|c|c|c|c|c|c|c|}{12.5cm}{Continued}
{\hline
OGLE-IV name & OGLE-IV field & RA & DEC & $R_{\rm KDE}$ [\arcs] & name & cluster type \\ 
\hline
OGLE-MBR-CL-0061 & MBR109.08 &  2\uph14\upm38\zdot\ups91 & $-74\arcd21\arcm30\zdot\arcs4$ & 22 & BSBD4 & C \\ 
OGLE-MBR-CL-0062 & MBR109.11 &  2\uph08\upm13\zdot\ups12 & $-74\arcd31\arcm48\zdot\arcs3$ & 34 & WG17 & A \\ 
OGLE-MBR-CL-0063 & MBR109.11 &  2\uph07\upm47\zdot\ups97 & $-74\arcd26\arcm31\zdot\arcs8$ & 26 & BS229 & C \\ 
OGLE-MBR-CL-0064 & MBR109.11 &  2\uph07\upm40\zdot\ups03 & $-74\arcd37\arcm47\zdot\arcs1$ & 34 & WG15 & C \\ 
OGLE-MBR-CL-0065 & MBR109.12 &  2\uph05\upm40\zdot\ups86 & $-74\arcd23\arcm00\zdot\arcs1$ & 36 & BS226 & C \\ 
OGLE-MBR-CL-0066 & MBR109.13 &  2\uph04\upm45\zdot\ups46 & $-74\arcd30\arcm57\zdot\arcs6$ & 22 & WG14 & C \\ 
OGLE-MBR-CL-0067 & MBR109.13 &  2\uph04\upm02\zdot\ups81 & $-74\arcd28\arcm46\zdot\arcs9$ & 38 & BS223 & C \\ 
OGLE-MBR-CL-0068 & MBR109.14 &  2\uph00\upm38\zdot\ups09 & $-74\arcd33\arcm30\zdot\arcs8$ & 20 & WG11 & C \\ 
OGLE-MBR-CL-0069 & MBR109.15 &  1\uph59\upm59\zdot\ups09 & $-74\arcd22\arcm57\zdot\arcs5$ & 45 & WG10 & AC \\ 
OGLE-MBR-CL-0070 & MBR109.18 &  2\uph11\upm49\zdot\ups50 & $-74\arcd06\arcm59\zdot\arcs0$ & 31 & BS235 & C \\ 
OGLE-MBR-CL-0071 & MBR109.19 &  2\uph10\upm40\zdot\ups97 & $-74\arcd09\arcm20\zdot\arcs6$ & 27 & BS233 & CA \\ 
OGLE-MBR-CL-0072 & MBR109.19 &  2\uph11\upm12\zdot\ups23 & $-74\arcd16\arcm44\zdot\arcs9$ & 21 & BS234 & AC \\ 
OGLE-MBR-CL-0073 & MBR109.24 &  1\uph59\upm47\zdot\ups87 & $-74\arcd16\arcm30\zdot\arcs4$ & 34 & BS220 & A \\ 
OGLE-MBR-CL-0074 & MBR109.25 &  1\uph56\upm55\zdot\ups40 & $-74\arcd15\arcm20\zdot\arcs6$ & 30 & BS216/BS217 & C/A \\ 
OGLE-MBR-CL-0075 & MBR109.25 &  1\uph56\upm44\zdot\ups78 & $-74\arcd13\arcm08\zdot\arcs1$ & 24 & NGC796,L115,WG9,ESO30SC6/BS215 & CA/A \\ 
OGLE-MBR-CL-0076 & MBR109.25 &  1\uph56\upm35\zdot\ups44 & $-74\arcd16\arcm58\zdot\arcs3$ & 25 & WG8 & AC \\ 
OGLE-MBR-CL-0077 & MBR109.28 &  2\uph09\upm20\zdot\ups82 & $-74\arcd01\arcm38\zdot\arcs3$ & 24 & BS232/BS231 & CA/A \\ 
OGLE-MBR-CL-0078 & MBR109.30 &  2\uph02\upm44\zdot\ups28 & $-73\arcd56\arcm16\zdot\arcs0$ & 22 & WG13 & C \\ 
OGLE-MBR-CL-0079 & MBR110.30 &  2\uph04\upm21\zdot\ups20 & $-74\arcd59\arcm01\zdot\arcs8$ & 34 & ICA6 & A \\ 
OGLE-MBR-CL-0080 & MBR113.06 &  2\uph19\upm28\zdot\ups70 & $-74\arcd11\arcm45\zdot\arcs4$ & 31 & BS243 & A \\ 
OGLE-MBR-CL-0081 & MBR113.08 &  2\uph31\upm11\zdot\ups50 & $-73\arcd55\arcm51\zdot\arcs0$ & 28 & ICA57 & A \\ 
OGLE-MBR-CL-0082 & MBR113.10 &  2\uph27\upm16\zdot\ups01 & $-73\arcd45\arcm38\zdot\arcs6$ & 40 & IDK2w,ICA45 & A \\ 
OGLE-MBR-CL-0083 & MBR113.10 &  2\uph27\upm28\zdot\ups38 & $-73\arcd58\arcm29\zdot\arcs4$ & 31 & BS245 & CA \\ 
OGLE-MBR-CL-0084 & MBR113.10 &  2\uph28\upm22\zdot\ups51 & $-73\arcd48\arcm05\zdot\arcs4$ & 28 & ICA49/ICA48 & A/A \\ 
OGLE-MBR-CL-0085 & MBR113.16 &  2\uph14\upm50\zdot\ups33 & $-73\arcd57\arcm10\zdot\arcs9$ & 31 & BS240/ICA34 & C/A \\ 
\hline}
\setcounter{table}{1}
\MakeTableSepp{|c|c|c|c|c|c|c|}{12.5cm}{Continued}
{\hline
OGLE-IV name & OGLE-IV field & RA & DEC & $R_{\rm KDE}$ [\arcs] & name & cluster type \\ 
\hline
OGLE-MBR-CL-0086 & MBR113.16 &  2\uph14\upm34\zdot\ups80 & $-73\arcd58\arcm56\zdot\arcs6$ & 28 & BS239/ICA34 & CA/A \\ 
OGLE-MBR-CL-0087 & MBR123.26 &  3\uph01\upm33\zdot\ups52 & $-73\arcd25\arcm08\zdot\arcs5$ & 24 & ICA71 & A \\ 
OGLE-MBR-CL-0088 & MBR128.03 &  3\uph10\upm22\zdot\ups86 & $-73\arcd30\arcm07\zdot\arcs5$ & 18 & BS247 & AC \\ 
OGLE-MBR-CL-0089 & MBR128.15 &  3\uph01\upm33\zdot\ups27 & $-73\arcd25\arcm08\zdot\arcs2$ & 24 & ICA71 & A \\ 
OGLE-MBR-CL-0090 & MBR141.07 &  3\uph44\upm26\zdot\ups41 & $-71\arcd40\arcm50\zdot\arcs2$ & 47 & NGC1466,SL1,LW1,ESO54SC16,KMHK1 & C \\ 
OGLE-MBR-CL-0091 & MBR160.11 &  1\uph55\upm36\zdot\ups02 & $-77\arcd39\arcm15\zdot\arcs5$ & 17 & L116,ESO13SC25 & C \\ 
OGLE-SMC-CL-0279 & SMC738.06 &  1\uph29\upm27\zdot\ups77 & $-73\arcd31\arcm56\zdot\arcs5$ & 27 & B164 & C \\ 
OGLE-SMC-CL-0280 & SMC738.06 &  1\uph29\upm34\zdot\ups82 & $-73\arcd33\arcm29\zdot\arcs4$ & 34 & GHK24/GHK29/GKH22/GHK51/NGC602,L105,ESO29SC43,H-A68 & C/C/C/C/DAN \\ 
OGLE-SMC-CL-0281 & SMC738.06 &  1\uph29\upm14\zdot\ups50 & $-73\arcd32\arcm02\zdot\arcs1$ & 36 & SGDH-cluster-A & C \\ 
OGLE-SMC-CL-0282 & SMC738.08 &  1\uph42\upm53\zdot\ups34 & $-73\arcd20\arcm15\zdot\arcs3$ & 25 & WG1 & C \\ 
OGLE-SMC-CL-0283 & SMC738.12 &  1\uph34\upm41\zdot\ups19 & $-73\arcd16\arcm27\zdot\arcs2$ & 29 & H86-213 & C \\ 
OGLE-SMC-CL-0284 & SMC738.13 &  1\uph31\upm08\zdot\ups83 & $-73\arcd24\arcm51\zdot\arcs1$ & 41 & L107,H-A69 & AC \\ 
OGLE-SMC-CL-0285 & SMC738.13 &  1\uph30\upm49\zdot\ups89 & $-73\arcd25\arcm45\zdot\arcs2$ & 43 & B165 & C \\ 
OGLE-SMC-CL-0286 & SMC738.13 &  1\uph30\upm33\zdot\ups40 & $-73\arcd25\arcm20\zdot\arcs7$ & 46 & BS186 & A \\ 
OGLE-SMC-CL-0287 & SMC738.16 &  1\uph25\upm25\zdot\ups86 & $-73\arcd22\arcm58\zdot\arcs1$ & 46 & BS282/L104/H-A67 & C/AN/DAN \\ 
OGLE-SMC-CL-0288 & SMC738.16 &  1\uph24\upm30\zdot\ups28 & $-73\arcd24\arcm41\zdot\arcs9$ & 46 & H86-211 & C \\ 
OGLE-SMC-CL-0289 & SMC738.16 &  1\uph24\upm09\zdot\ups76 & $-73\arcd09\arcm27\zdot\arcs2$ & 62 & HW81 & CN \\ 
OGLE-SMC-CL-0290 & SMC738.16 &  1\uph24\upm25\zdot\ups25 & $-73\arcd10\arcm31\zdot\arcs2$ & 46 & HW82 & C \\ 
OGLE-SMC-CL-0291 & SMC738.16 &  1\uph24\upm25\zdot\ups37 & $-73\arcd10\arcm30\zdot\arcs4$ & 57 & BS176/HCD99-1 & C/C \\ 
OGLE-SMC-CL-0292 & SMC738.21 &  1\uph34\upm25\zdot\ups67 & $-72\arcd52\arcm21\zdot\arcs8$ & 45 & L110,ESO29SC48 & C \\ 
OGLE-SMC-CL-0293 & SMC738.22 &  1\uph31\upm01\zdot\ups36 & $-72\arcd51\arcm03\zdot\arcs1$ & 28 & BS187 & CA \\ 
OGLE-SMC-CL-0294 & SMC739.20 &  1\uph33\upm12\zdot\ups46 & $-74\arcd10\arcm02\zdot\arcs7$ & 24 & L109,ESO29SC46 & C \\ 
OGLE-SMC-CL-0295 & SMC739.29 &  1\uph31\upm11\zdot\ups93 & $-73\arcd53\arcm35\zdot\arcs6$ & 45 & B166 & C \\ 
OGLE-SMC-CL-0296 & SMC740.03 &  1\uph30\upm38\zdot\ups30 & $-76\arcd03\arcm15\zdot\arcs3$ & 28 & L106,ESO29SC44 & C \\ 
OGLE-SMC-CL-0297 & SMC740.18 &  1\uph34\upm55\zdot\ups99 & $-75\arcd33\arcm17\zdot\arcs1$ & 38 & NGC643,L111,ESO29SC50 & C \\ 
\hline}
\setcounter{table}{1}
\MakeTableSepp{|c|c|c|c|c|c|c|}{12.5cm}{Continued}
{\hline
OGLE-IV name & OGLE-IV field & RA & DEC & $R_{\rm KDE}$ [\arcs] & name & cluster type \\ 
\hline
OGLE-SMC-CL-0298 & SMC740.18 &  1\uph35\upm58\zdot\ups27 & $-75\arcd27\arcm26\zdot\arcs0$ & 23 & L112 & C \\ 
OGLE-SMC-CL-0299 & SMC740.31 &  1\uph22\upm44\zdot\ups71 & $-75\arcd00\arcm30\zdot\arcs4$ & 39 & HW79 & C \\ 
OGLE-SMC-CL-0300 & SMC737.14 &  1\uph31\upm38\zdot\ups99 & $-71\arcd56\arcm49\zdot\arcs4$ & 29 & L108 & C \\ 
OGLE-SMC-CL-0301 & SMC737.17 &  1\uph43\upm52\zdot\ups43 & $-71\arcd44\arcm51\zdot\arcs9$ & 31 & BS190 & CA \\ 
OGLE-SMC-CL-0302 & SMC737.21 &  1\uph35\upm11\zdot\ups61 & $-71\arcd44\arcm15\zdot\arcs5$ & 37 & BS188 & C \\ 
OGLE-SMC-CL-0303 & SMC737.32 &  1\uph30\upm11\zdot\ups16 & $-71\arcd20\arcm19\zdot\arcs7$ & 27 & BS184 & CA \\ 
OGLE-SMC-CL-0304 & SMC736.01 &  1\uph42\upm27\zdot\ups76 & $-71\arcd16\arcm47\zdot\arcs8$ & 20 & HW85 & C \\ 
OGLE-SMC-CL-0305 & SMC736.02 &  1\uph41\upm40\zdot\ups48 & $-71\arcd09\arcm53\zdot\arcs2$ & 30 & HW84 & C \\ 
OGLE-SMC-CL-0306 & SMC734.08 &  1\uph22\upm49\zdot\ups14 & $-75\arcd00\arcm04\zdot\arcs9$ & 41 & HW79 & C \\ 
OGLE-SMC-CL-0307 & SMC734.12 &  1\uph12\upm04\zdot\ups82 & $-75\arcd11\arcm40\zdot\arcs1$ & 38 & HW66,ESO29SC36 & C \\ 
OGLE-SMC-CL-0308 & SMC717.25 &  0\uph48\upm50\zdot\ups91 & $-69\arcd52\arcm08\zdot\arcs7$ & 40 & L38,ESO51SC3 & C \\ 
OGLE-SMC-CL-0309 & SMC716.10 &  0\uph58\upm58\zdot\ups01 & $-68\arcd54\arcm54\zdot\arcs0$ & 23 & ESO51SC9 & C \\ 
OGLE-SMC-CL-0310 & SMC710.26 &  0\uph47\upm24\zdot\ups56 & $-68\arcd55\arcm15\zdot\arcs1$ & 25 & L32,ESO51SC2 & C \\ 
OGLE-SMC-CL-0311 & SMC706.12 &  0\uph26\upm52\zdot\ups99 & $-71\arcd32\arcm56\zdot\arcs6$ & 46 & NGC121,K2,L10,ESO50SC12 & C \\ 
OGLE-SMC-CL-0312 & SMC703.01 &  0\uph12\upm57\zdot\ups34 & $-73\arcd29\arcm30\zdot\arcs6$ & 30 & L2 & C \\ 
OGLE-SMC-CL-0313 & SMC703.05 &  0\uph03\upm47\zdot\ups83 & $-73\arcd28\arcm43\zdot\arcs4$ & 24 & L1,ESO28SC8 & C \\ 
OGLE-SMC-CL-0314 & SMC715.28 &  0\uph22\upm42\zdot\ups73 & $-75\arcd04\arcm33\zdot\arcs8$ & 23 & L5,ESO28SC16 & C \\ 
OGLE-SMC-CL-0315 & SMC761.02 & 23\uph48\upm59\zdot\ups38 & $-72\arcd56\arcm43\zdot\arcs6$ & 16 & AM-3,ESO28SC4 & C \\ 
OGLE-SMC-CL-0316 & SMC707.01 &  0\uph28\upm31\zdot\ups18 & $-73\arcd00\arcm40\zdot\arcs4$ & 50 & BS2 & C \\ 
OGLE-SMC-CL-0317 & SMC707.03 &  0\uph24\upm57\zdot\ups18 & $-73\arcd01\arcm48\zdot\arcs4$ & 40 & B4 & CA \\ 
OGLE-SMC-CL-0318 & SMC707.09 &  0\uph27\upm44\zdot\ups16 & $-72\arcd46\arcm46\zdot\arcs9$ & 46 & K7,L11,ESO28SC22 & C \\ 
OGLE-SMC-CL-0319 & SMC707.11 &  0\uph24\upm44\zdot\ups77 & $-72\arcd47\arcm45\zdot\arcs0$ & 50 & K3,L8,ESO28SC19 & C \\ 
OGLE-SMC-CL-0320 & SMC707.17 &  0\uph31\upm03\zdot\ups58 & $-72\arcd20\arcm21\zdot\arcs1$ & 37 & HW5 & C \\ 
OGLE-SMC-CL-0321 & SMC707.29 &  0\uph21\upm30\zdot\ups47 & $-71\arcd56\arcm03\zdot\arcs5$ & 35 & BOLOGNA-A & C \\ 
\hline}
\setcounter{table}{1}
\MakeTableSepp{|c|c|c|c|c|c|c|}{12.5cm}{Concluded}
{\hline
OGLE-IV name & OGLE-IV field & RA & DEC & $R_{\rm KDE}$ [\arcs] & name & cluster type \\ 
\hline
OGLE-SMC-CL-0322 & SMC708.03 &  0\uph19\upm19\zdot\ups65 & $-74\arcd06\arcm23\zdot\arcs3$ & 36 & B1 & C \\ 
OGLE-SMC-CL-0323 & SMC708.04 &  0\uph18\upm25\zdot\ups79 & $-74\arcd19\arcm07\zdot\arcs0$ & 22 & L3,ESO28SC13 & C \\ 
OGLE-SMC-CL-0324 & SMC714.31 &  0\uph24\upm39\zdot\ups57 & $-73\arcd45\arcm11\zdot\arcs9$ & 45 & K5,L7,ESO28SC18 & C \\ 
OGLE-SMC-CL-0325 & SMC708.10 &  0\uph21\upm31\zdot\ups25 & $-73\arcd45\arcm27\zdot\arcs1$ & 45 & K1,L4,ESO28SC15 & C \\ 
OGLE-SMC-CL-0326 & SMC708.18 &  0\uph23\upm03\zdot\ups83 & $-73\arcd40\arcm09\zdot\arcs5$ & 37 & K4,L6,ESO28SC17 & C \\ 
OGLE-SMC-CL-0327 & SMC708.19 &  0\uph21\upm27\zdot\ups97 & $-73\arcd44\arcm54\zdot\arcs1$ & 41 & K1,L4,ESO28SC15 & C \\ 
OGLE-SMC-CL-0328 & SMC708.23 &  0\uph12\upm55\zdot\ups25 & $-73\arcd29\arcm27\zdot\arcs9$ & 29 & L2 & C \\ 
OGLE-SMC-CL-0329 & SMC708.29 &  0\uph18\upm23\zdot\ups44 & $-73\arcd23\arcm40\zdot\arcs5$ & 36 & HW1 & CA \\ 
OGLE-SMC-CL-0330 & SMC714.12 &  0\uph28\upm39\zdot\ups66 & $-74\arcd23\arcm55\zdot\arcs6$ & 51 & B6 & C \\ 
OGLE-SMC-CL-0331 & SMC714.16 &  0\uph19\upm18\zdot\ups10 & $-74\arcd34\arcm26\zdot\arcs2$ & 25 & B2 & C \\ 
OGLE-SMC-CL-0332 & SMC714.22 &  0\uph25\upm26\zdot\ups81 & $-74\arcd04\arcm30\zdot\arcs9$ & 40 & K6,L9,ESO28SC20 & C \\ 
OGLE-SMC-CL-0333 & SMC724.03 &  1\uph10\upm43\zdot\ups92 & $-71\arcd16\arcm50\zdot\arcs2$ & 51 & BS144 & A \\ 
OGLE-SMC-CL-0334 & SMC724.07 &  1\uph02\upm01\zdot\ups10 & $-71\arcd01\arcm11\zdot\arcs5$ & 40 & B111 & C \\ 
OGLE-SMC-CL-0335 & SMC724.09 &  1\uph13\upm03\zdot\ups80 & $-70\arcd57\arcm46\zdot\arcs1$ & 35 & HW67 & C \\ 
OGLE-SMC-CL-0336 & SMC724.12 &  1\uph07\upm41\zdot\ups73 & $-70\arcd56\arcm08\zdot\arcs4$ & 26 & HW56 & C \\ 
OGLE-SMC-CL-0337 & SMC724.31 &  1\uph04\upm24\zdot\ups97 & $-70\arcd20\arcm32\zdot\arcs3$ & 26 & L73 & C \\ 
OGLE-SMC-CL-0338 & SMC731.08 &  1\uph30\upm11\zdot\ups68 & $-71\arcd20\arcm17\zdot\arcs5$ & 33 & BS184 & CA \\ 
OGLE-SMC-CL-0339 & SMC731.15 &  1\uph16\upm24\zdot\ups75 & $-71\arcd19\arcm36\zdot\arcs1$ & 33 & HW73 & C \\ 
OGLE-SMC-CL-0340 & SMC731.16 &  1\uph14\upm54\zdot\ups34 & $-71\arcd32\arcm32\zdot\arcs6$ & 48 & NGC458,K69,L96,ESO51SC26 & C \\ 
OGLE-SMC-CL-0341 & SMC731.16 &  1\uph14\upm44\zdot\ups48 & $-71\arcd20\arcm54\zdot\arcs3$ & 38 & L95 & C \\ 
OGLE-SMC-CL-0342 & SMC731.20 &  1\uph24\upm55\zdot\ups87 & $-71\arcd11\arcm13\zdot\arcs3$ & 26 & IC1708,L102,ESO52SC2 & C \\ 
OGLE-SMC-CL-0343 & SMC731.27 &  1\uph26\upm42\zdot\ups70 & $-70\arcd46\arcm58\zdot\arcs8$ & 24 & B168 & C \\ 
OGLE-SMC-CL-0344 & SMC739.05 &  1\uph29\upm52\zdot\ups83 & $-74\arcd50\arcm48\zdot\arcs2$ & 28 & Field12-01 & - \\ 
OGLE-SMC-CL-0345 & SMC734.21 &  1\uph13\upm42\zdot\ups75 & $-74\arcd45\arcm14\zdot\arcs4$ & 28 & Field16-01 & - \\ 
\hline}
\end{landscape}

\renewcommand{\arraystretch}{1}
\renewcommand{\TableFont}{\footnotesize}
\MakeTable{|c|c|c|}{12.5cm}{Undetected objects form Bica \etal (2008) and Piatti (2017)}
{\hline
\douprule
Name & Bica type & Comment \\
\hline
BS6        & CA & not visible \\
H86-197    & C  & edge \\
BS127      & CA &  not visible \\
HW20       & C  & edge \\
B44        & C  & edge \\
BS173      & CA & not visible \\
BS1        & CA & not visible \\
BS189      & CA & not visible \\
Field16-02 & -- & edge \\
\hline}

The OGLE star cluster collection, the list of all analyzed SMC and MBR fields and all the graphical
materials are avaliable on the OGLE web page:\\ 
\centerline{\it  http://ogle.astrouw.edu.pl}

\Section{Conclusions}
We have presented a catalog of star clusters in the Magellanic Bridge
and the outer regions of the Small Magellanic Cloud based on the
OGLE-IV deep photometric maps. We found a total of 198 star clusters,
including 75 new objects which were not listed in any of the previous
catalogs, 121 clusters listed in Bica \etal (2008) and two clusters
listed in Piatti (2017). For all of them the equatorial coordinates
and cross-identification with the Bica \etal catalog are provided. The
detection method presented in this paper is very effective. With our
algorithm we found more than 95\% of previously known clusters in this
characteristic sparse region of the SMC and in the whole MBR,
increasing the total number of these objects by 40\%. This paper is
the second of a series of publications. In the next one we will
present clusters found in the central regions of the LMC and SMC, thus
concluding the complete collection of star clusters in the whole
Magellanic System observed by the OGLE survey.

\Acknow{We gratefully acknowledge the financial support of the Polish
  National Science Center, grant SONATA 2013/11/D/ST9/03445 to D.~Skow\-ron.
  Z.~Kostrzewa-Rutkowska acknowledges support from European Research
  Council Consolidator Grant 647208. The OGLE project has received funding
  from the National Science Center, Poland, grant MAESTRO
  2014/14/A/ST9/00121 to A.~Udalski.}

\end{document}